\documentstyle[prd,aps,preprint,eqsecnum,epsfig,floats]{revtex}
\title{ Nonperturbative hyperfine contribution to the $b_1$ \\
and $h_1$ meson masses.}
\author{A.M.Badalian$^{a}$ and B.L.G.Bakker$^b$}
\address{$^a$ Institute of Theoretical and Experimental Physics,\\
117218 Moscow, B.Cheremushkinskaya 25, Russia}
\address{$^b$ Department of Physics and Astronomy, Vrije Universiteit,
 Amsterdam, The Netherlands}
\date{DRAFT: \today}
\begin{document}

\maketitle

\newcommand{\be}{\begin{equation}}
\newcommand{\ee}{\end{equation}}

\def\la{\mathrel{\mathpalette\fun <}}
\def\ga{\mathrel{\mathpalette\fun >}}
\def\fun#1#2{\lower3.6pt\vbox{\baselineskip0pt\lineskip.9pt
\ialign{$\mathsurround=0pt#1\hfil##\hfil$\crcr#2\crcr\sim\crcr}}}

\vspace{1cm}

\begin{abstract}
Due to the nonperturbative contribution to the hyperfine splitting the mass
of the $n^1P_1$ state is strongly correlated with  the center of gravity
$M_{\rm cog}(n^3P_J)$  of the $n^3P_J$  multiplet: $M(n^1P_1)$ is less than
$M_{\rm cog}(n^3P_J)$  by about 40 MeV (20 MeV)  for the 1P (2P)  state. For
$b_1(1235)$ the agreement with experiment is reached only if $a_0(980)$
belongs to the $1^3P_J$  multiplet. The predicted mass of $b_1(2^1P_1)$ is
$\approx 1620$  MeV. For the isoscalar meson a correlation between the mass of
$h_1$(1170) $(h_1(1380))$ and $M_{\rm cog}(1^3P_J)$  composed from light
(strange) quarks also takes place.

\end{abstract}

\section{Introduction}

Since the discovery of the $h_c$  meson \cite{ref.01} the hyperfine
(HF)  splittings of the $P$-wave states in heavy quarkonia were
investigated in many papers \cite{ref.02}-\cite{ref.06}. In Refs.~
\cite{ref.05}-\cite{ref.06}  it was clarified why the HF shift of
the $h_c$ meson with respect to the center of gravity $M_{\rm cog}(^3P_J)$
of the $\chi_c$  mesons turns out to be small, $\Delta_{\rm
HF}(h_c)=-0.87\pm 0.24$ MeV \cite{ref.07}. It is  due to a cancellation of
the perturbative and nonperturbative contributions which are both small and
have  opposite signs:  $\Delta^{\rm P}_{\rm HF}(c\bar{c})\approx
-1.7\pm 0.3$  MeV and $\Delta^{\rm NP}_{\rm HF}(c\bar{c})\approx
1$ MeV. Here the total HF shift $\Delta_{\rm HF}$ is defined in the following way
\begin{equation}
 \Delta_{\rm HF} = M_{\rm cog} (n{}^3 P_J) - M(n{}^1 P_1).
 \label{eq.001}
\end{equation}

For light mesons the HF splittings of the $P$-wave states are of special
interest, since for them the perturbative spin-spin interaction is
suppressed as for any $L=1$ state, while the nonperturbative HF
interaction is expected to become larger. In our study it will be shown
that the nonperturbative contribution $\Delta^{\rm NP}_{\rm HF}$,
defined through the vacuum correlators, does dominate and $\Delta_{\rm
HF}$(1P) is about 30 MeV.  Although the magnitude of the splitting
depends on such vacuum characteristics as the gluon condensate $G_2$
and the gluonic correlation length $T_g$, the total $\Delta_{\rm
HF}(nP)$ turns out to be positive in all cases considered.

In our calculations of the HF splittings we shall follow the approach
developed in Ref.~\cite{ref.08}  where the spin-dependent interaction is
considered as a perturbation and averaging the spin factors in a meson
Green's function is performed without the expansion in inverse powers
of quark masses, used in the usual treatment \cite{ref.09}.  Therefore
the spin-spin potential from Ref.~\cite{ref.08}  can be used for
massless quarks and the HF splittings appear to be proportional to
$[\mu_0(nL)]^{-2}$ where $\mu_0(nL)$ is the effective dynamical mass of a
light quark which is defined by the extremum of the Hamiltonian deduced
from the QCD Lagrangian.  It is essential that $\mu_0(nL)$ depends on
the quantum numbers of the state considered and is not small; for the
$nP$ meson containing a light quark and antiquark, $\mu_0(1P)\approx 0.40$  GeV
and $\mu_0(2P)\approx 0.52$ GeV and $\mu_0(1P)=454$ MeV,
$\mu_0(2P)=566$ MeV for the $nP\, s\bar{s}$ states.

For the isovector $1P$ mesons $(b_1(1235)$ and the ground states of the
$a_J$ mesons) the calculated $\Delta_{\rm HF}(1P)$ is 39(19) MeV for
two different vacuum gluonic correlation lengths: $T_g = 0.3(0.2)$ fm, and
with the use of the experimental mass of $b_1(1235)$ we obtain that
\be
M_{\rm cog}(1^3P_J,\, I=1) = 1258 \pm 10 \; \mbox{MeV},
\label{eq.01}
\ee
where the theoretical error comes from the uncertainty in the value of
the gluonic length $T_g$.
From this result an important consequence follows, namely, the
number (\ref{eq.01}) is compatible with the experimental  masses of the
$a_J$  mesons $(n=1)$ only if $a_0(980)$  (but not $a_0(1450)$)
belongs to the isovector $1^3P_J$  multiplet, i.e. $a_0(980)$  is a
usual $q\bar{q}$  state.

For the $b_1(2P)$  meson the mass $M(b_1(2P))\approx 1620$ MeV is
predicted.  The situation with the isoscalar $P$-wave mesons  ($h_1$
and $f_J$)  is also discussed and a correlation between the masses of
$h_1(1170)$ and $M_{\rm cog}(1^3P_J)$=1245 MeV for
$f_0(980),\, f_1(1285)$, $f_2(1270)$, as well as between the mass of
$h_1(1380)$ and $M_{\rm cog}(1^3P_J)\approx 1420$ MeV for $f_0(1370)$,
$f_1(1420)$, $f_2(1430)$ (or $M_{\rm cog} = 1470$ MeV if $f'_2(1525)$
belongs to a multiplet composed of a strange quark and antiquark)
can also be interpreted as a manifestation of
a positive $(\approx 30$ MeV) nonperturbative HF splitting.

\section{Nonperturbative hyperfine interaction}

The HF splitting of the $P$-wave mesons originates both from
perturbative and nonperturbative interactions:
\be
\Delta_{\rm HF}(nP) = \Delta^{\rm P}_{\rm HF} \, (nP) +
\Delta_{\rm HF}^{\rm NP} (nP) ,
\label{eq.13}
\ee
where the perturbative term for $L=1$  exists only in second order of
$\alpha_s$ and will be discussed in Sec. 5. The quantity $\Delta^{\rm NP}_{\rm
HF}$ is defined by the nonperturbative spin-spin potential which is
usually presented in the form,
\be
V^{\rm NP}_{\rm HF} (r) = \frac{1}{3m^2_q} \, V^{\rm NP}_4 (r) .
\label{eq.14}
\ee
As was shown in Ref.~\cite{ref.08} the spin-spin potential  $V^{\rm
NP}_4 (r)$ appears to be the same for heavy and light mesons (if the
spin-dependent interaction is considered as a perturbation) and can
be expressed through the vacuum correlators $D(x)$ and $D_1(x)$ which
were introduced in Ref.~\cite{ref.10} and calculated in lattice QCD
\cite{ref.11}-\cite{ref.12}
\be
V^{\rm NP}_4(r) = 2 \; \int\limits^{\infty}_0 \, d\nu \Biggl [3D(r,\nu) +
3D_1(r,\nu) + 2r^2 \frac{\partial D_1(r,\nu)}{\partial r^2}\Biggr ].
\label{eq.15}
\ee
By definition, at the origin $(x=0)$  these correlators are related to
the gluon condensate $G_2=\frac{\alpha_s}{\pi} \langle
F^a_{\mu\nu}(0)F^a_{\mu\nu}(0)\rangle $:
\be
D(0) + D_1(0) = \frac{\pi^2}{18}G_2,
\label{eq.16}
\ee
where  the physical value of $G_2=0.04 \pm 0.02$ GeV$^4$  is usually
taken.

In lattice calculations it was found that  $D(x)$  and $D_1(x)$  can be
parametrized as exponentials at separations $x \ga 0.2$ fm:
\cite{ref.11}-\cite{ref.13}
\be
D(x) = d \, \exp \left( -\frac{x}{T_g} \right), \quad
D_1(x) = d_1 \, \exp \left( -\frac{x}{T^{(1)}_g} \right),\quad
(x > 0.2\; \mbox{fm}),
\label{eq.17}
\ee
with the gluonic correlation lengths $T_g$ and $T_g^{(1)}$  which
turn out to be different in the quenched approximation and full QCD. In the
general case the parameters $d$ and $d_1$, obtained in lattice
measurements, differ from $D(0)$ and $D_1(0)$.

\underline{In full QCD} with dynamical fermions ($n_f=4$) the
correlation length was found to be relatively large and the
$D_1$--correlator is small and can be neglected in some cases\cite{ref.12}:
\be
T_g \approx 0.3\; \mbox{fm}, \quad d_1 \approx \frac{1}{10}d, \quad (n_f=4).
\label{eq.18}
\ee
It was shown in Ref.~\cite{ref.12} that in this case the correlator $D(x)$ can be taken as
an exponential over all distances, i.e. $d=D(0)$,
\be
D(x) = D(0) \, \exp \left(-\frac{x}{T_g} \right), \quad (T_g\approx 0.3\, \mbox{fm})
\label{eq.19}
\ee
and from Eq.~(\ref{eq.16}) in this case
\be
D(0) \approx \frac{\pi^2}{18}\, G_2 = 0.55 \; G_2.
\label{eq.20}
\ee
Then from Eq.~(\ref{eq.15})  the potential $V_4^{\rm NP}(r)$ is given by
the expression:
\be
V^{\rm NP}_4 (r) = 6\,d\;\int\limits^{\infty}_0 \exp \Biggl
(-\frac{\sqrt{r^2+\nu^2}}{T_g}\Biggr )d\nu
 = 6\, d\,r\,K_1 \left(\frac{r}{T_g} \right), \quad d= D(0) .
\label{eq.21}
\ee
The string tension $\sigma$ is defined in the general case as
\be
\sigma = 2 \int\limits^{\infty}_0 d\nu\int\limits^{\infty}_0 d\lambda
\,D(\sqrt{\lambda^2 + \nu^2}) ,
\label{eq.22}
\ee
and for $D(x)$  taken as an exponential at all distances it reduces to
the relation
\be
\sigma = \pi d \, T_g^2 \quad \mbox{or} \quad d = \frac{\sigma}{\pi T_g^2},
 \quad G_2 \approx \frac{18\sigma}{\pi^3 T^2_g} .
\label{eq.23}
\ee
If $\sigma$  is fixed and not large $(\sigma\approx 0.14$ GeV$^2$)  then
for the gluon condensate a reasonable value $0.036$ GeV$^4$ (for
$T_g=0.3$ fm) follows.  In this case the nonperturbative HF
splitting is
\be
\Delta^{\rm NP}_{\rm HF} (nP) = \frac{2d}{m^2_q} \langle r K_1 (r/T_g)
\rangle_{nP} = 
 \frac{2\sigma}{\pi T_g^2 m_q^2}  \langle r K_1 (r/T_g) \rangle_{nP} .
\label{eq.25}
\ee
For light mesons the HF shift in the form of the relation
(\ref{eq.25})  gives a dominant contribution also in cases when $D(x)$
cannot be interpolated up to the origin, see below. The matrix elements
in Eq.~(\ref{eq.25}) will be calculated in our paper with the use of
the solutions of the spinless Salpeter equation and the definition of
the effective mass $m_q$ of a light quark  will be discussed in the
next section.

Here we would like to notice that the potential $V^{\rm NP}_4(r)$  in
Eq.~(\ref{eq.21}), corresponding to the exponential correlator from
Ref.~\cite{ref.12}, has an essential shortcoming. From our calculations
it follows that this term gives a rather large nonperturbative shift in 
charmonium,
\be
\Delta^{\rm NP}_{\rm HF} (1P,\, c\bar{c}) \ga 5.0\,\mbox{MeV}, \quad (T_g =0.3
 \, \mbox{fm}) ,
\label{eq.26}
\ee
so that the total splitting (\ref{eq.13}) turns out to be positive for
$h_c$ in contradiction with the  experimental negative number.
Therefore, to explain the HF splitting of the 1P state in charmonium one
needs to know $D(x)$ in detail at small distances, since the HF
splitting in heavy quarkonia appears to be very sensitive to the
behavior of the correlators $D(x)$ and $D_1(x)$ at short distances (this
problem will be considered in another paper).  However,  for the
light $P$-wave mesons the behavior of the correlators $D(x)$   and
$D_1(x)$ at  short distances was found to be inessential and for them
the potential $V^{\rm NP}_4(r)$ in the form of Eq.~(\ref{eq.21}) can be
used with 5-10\% accuracy.

Nevertheless, for completeness  we give below expressions for the
correlator $D(x)$ and for $V^{\rm NP}_4(r)$, modified such as to
make clear that there exists the opportunity to combine a small, ``physical''
value of the gluonic condensate $G_2$ and a small correlation length
$T_g$. Otherwise the values fitted in lattice calculations (quenched
approximation), $T_g\approx 0.2$ fm in Ref.~\cite{ref.11}  and
$T_g\approx 0.12$  fm in Ref.~\cite{ref.13}, give rise to very large
``unphysical'' values of $G_2$, $\approx 0.14$  GeV$^4$  and $0.23$
GeV$^4$, respectively.

To this end $D(x)$ is supposed to be a constant at $x < x_0$ which differs
from from the coefficient $d$ in Eq.~(\ref{eq.17}) and can be taken as
\be
D(x) = {\rm constant} = d\,\exp \left(-\frac{x_0}{T_g}\right), \quad x \la x_0,
\quad  x_0\approx 0.2 \; {\rm fm},
\label{eq.27}
\ee
while at $x \geq x_0$   $D(x)$ is given by the exponential (\ref{eq.19})
as it was observed in lattice measurements. Then even for very small
$T_g=0.6$ GeV$^{-1} = 0.12$ fm the small value $G_2\approx 0.02$ GeV$^4$
can be obtained for the gluon condensate. For the modified correlator
$D(x)$, Eq.~ (\ref{eq.27}), the modified nonperturbative spin-spin potential is
\begin{eqnarray}
\tilde{V}^{\rm NP}_4 (r) & = & 6d \Biggl [ e^{-\frac{x_0}{T_g}}
 \sqrt{x_0^2 - r^2} +
\int\limits^{\infty}_{\sqrt{x^2_0-r^2}} \, d\nu \, exp \Biggl (
-\frac{\sqrt{r^2+\nu^2}}{T_g} \Biggr ) \Biggr ] \theta(x_0-r) \nonumber \\
 & & + 6d\, r\, K_1\left(\frac{r}{T_g} \right)\,\theta (r-x_0) .
\label{eq.28}
\end{eqnarray}

For the $P$-wave light mesons the difference in the nonperturbative HF
shift for the potential $V^{\rm NP}_4(r)$ and $\tilde{V}^{\rm NP}_4(r)$
does not exceed 10\% and therefore the simpler potential $V^{\rm
NP}_4(r)$, defined by Eq.~(\ref{eq.21}), can  be used.  Still for the
$h_c$ meson in charmonium such a modification of the spin-spin potential is 
important.

\section{Spectrum and matrix elements}

The fine structure and HF splittings in light mesons, with the
exception of $\pi$  and $K$, are typically much smaller than their
masses  and therefore the spin-dependent interaction can be considered
as a perturbation.  Then the choice of an unperturbed Hamiltonian is of
great importance and here the unperturbed  approximation is formulated
with the help of the spinless Salpeter equation,
\be
 \{ 2\,\sqrt{{\bf p}^2 +m^2} + V_0(r)\} \,\psi_{nL}(r) = E_{nL}\,\psi_{nL} (r),
 \label{eq.29}
\ee
where $m$ is the current mass of a quark and $V_0(r)$  is the static
potential. We have chosen this equation since under some assumptions it
can be deduced from the QCD Lagrangian. In particular, if in the
Feynmann--Schwinger representation \cite{ref.13,ref.14} the backward
trajectories are neglected, then for $L=0$ the  QCD Hamiltonian for the
spinless  quark (antiquark) coincides with  Eq.~(\ref{eq.29}) and for
$L=1$  the correction to the equation (\ref{eq.29}) is not large
\cite{ref.15}.  Therefore we can use the Salpeter equation for the
$P$-wave states.

For light mesons in Eq.~(\ref{eq.29}) the current mass is taken to be zero
and the static potential $V_0(r)$ is taken in the form of the Cornell potential,
\be
V_0(r) =-\frac{4}{3} \,\frac{\alpha_{\rm eff}}{r}  + \sigma r  + C_0 ,
\label{eq.30}
\ee
where $\alpha_{\rm eff}$ is an effective Coulomb constant. 
One can expect that for light mesons
which have the rather large size $R \ga 1$ fm, $(R=\sqrt{\langle
r^2\rangle})$, the value of  $\alpha_{\rm eff}$ will probably be close to the
so-called freezing value $\alpha_{\rm fr}=\alpha_{\rm eff}(r\to \infty)$
which  was found in Refs.~\cite{ref.16}, \cite{ref.17}, and has the value
\be
\alpha_{\rm fr} = 0.50 \pm 0.05,
\label{eq.31}
\ee
if the screening effects are neglected. However, even for such a large
$\alpha_{\rm eff}$, at long distances, $r \ga 6$ GeV$^{-1}$, the Coulomb
interaction is small compared to the linear confining potential
and in most cases can be neglected. Therefore we consider here two
variants:
\be
\alpha_{\rm eff} = 0   \quad ({\rm case A}), \quad 
\alpha_{\rm eff} = 0.45   \quad ({\rm case B}).
\label{eq.32}
\ee
To fix the string  tension $\sigma$ in the static potential
(\ref{eq.30})  one needs to take  into account that although the
Salpeter equation with a linear potential $\sigma r$ provides a linear
Regge trajectory, however, as shown on Refs.~\cite{ref.15}, the slope of the
Regge trajectory  for the Salpeter equation
\be
\alpha^{\prime} = \frac{1}{8\sigma}
\label{eq.33}
\ee
 differs from the slope $\alpha^{\prime}_{s t}$ in the string picture
 where
\be
 \alpha^{\prime}_{\rm st} = \frac{1}{2\pi \sigma_{\rm st}},
 \label{eq.34}
\ee
with the standard value of $\sigma_{\rm st} \approx 0.182$ GeV$^2$.
Therefore, to provide the experimentally observed slope the value of
$\sigma$ in the Salpeter equation should be taken smaller than $\sigma_{\rm
st}$:
\be
\sigma =\frac{\pi}{4} \,\sigma_{\rm st}  = 0.143\,\mbox{GeV}^2 .
\label{eq.35}
\ee
In most of our calculations just this number will be taken, but in some
cases the value $\sigma \approx \sigma_{\rm st}\approx 0.18$ GeV$^2$
will be also used for comparison. Thus in case A the static interaction
is characterized by the parameter $\sigma$ only, with its value given by
the number (\ref{eq.35}).  With this smaller value of $\sigma$ the
masses of the excited  states in our calculations will be lower than in
Ref.~\cite{ref.17} (where the same Salpeter equation was solved with
$\sigma_{\rm st}=0.18$ GeV$^2$) and closer to the experimental meson
masses for the excited states.

\section{Dynamical masses of light quarks}

In Refs.~\cite{ref.08} a relativistic Hamiltonian $H_R$ was derived
from the meson Green's function in the Feynman-Schwinger representation
with the use of the auxiliary field (einbein)  approach. For $L=0$ and
a spinless quark (antiquark) $H_R$ is given by the operator
\be
H_R = \frac{{\bf p}^2+m^2}{\mu(\tau)} + \mu(\tau)  + \frac{\sigma^2r^2}{2}
\int\limits^1_0 \frac{d\beta}{\nu(\beta)} + \frac{1}{2} \int\limits^1_0
\nu(\beta)d\beta
\label{eq.36}
\ee
where $\mu(\tau)$   and $\nu(\beta)$  are the auxiliary operators and
$\mu(\tau)$ is defined in the following way:
\be
 \mu(\tau)  = \frac{1}{2} \,\frac{dt}{d\tau} .
\label{eq.37}
\ee
In the definition (\ref{eq.37}) $\tau$  is the proper time and $t$  is
the actual time. With the use of the steepest descent method the
extremal values $\mu_{\rm ex}(\tau)=\mu_0$  and $\nu_{\rm ex}(\beta)=\nu_0$
can be obtained with the following result:
\be
 \mu_0 = \sqrt{{\bf p}^2 + m^2},\quad \nu_0 =\sigma r .
\label{eq.38}
\ee
%
Then the relativistic Hamiltonian $H_R$ in Eq.~(\ref{eq.36})
reduces to the spinless Salpeter operator
\be
\tilde{H}_R = \frac{{\bf p}^2 +m^2}{\mu_0} + \mu_0 + \sigma r \to
2\sqrt{{\bf p}+m^2}  + \sigma r
\label{eq.39}
\ee
In what follows the extremal value $\mu_0$, which is an operator, will
be replaced by the average of this operator which depends on the quantum
numbers $nL$ of the state considered, i.e.
\begin{eqnarray}
 \mu_0(nL) & = & \langle \sqrt{{\bf p}^2+m^2} \rangle_{nL}
 \quad \; \mbox{for} \; m\not= 0, \nonumber \\
 \mu_0(nL) & = & \langle \sqrt{{\bf p}^2} \rangle_{nL} 
 \quad \quad \quad \quad \mbox{if} \; m= 0,
\label{eq.40}
\end{eqnarray}
where $m$ is the current mass of a quark (antiquark)  and for light
quarks we take $m=0$, while for the strange quark $m_s=170$ MeV will
be used.

The definition (\ref{eq.40}) of the effective  mass of a light quark was already
discussed in Ref.~\cite{ref.18}  where it was shown that the
expectation value of $\tilde{H}_R$  in Eq.~(\ref{eq.39})   coincides
with that for the nonrelativistic Schr\"odinger Hamiltonian, if the
effective mass is defined as in Eq.~(\ref{eq.40}).

As seen from the definition (\ref{eq.40}) the dynamical mass of
a light quark $\mu_0(nL)$ appears to coincide with half the average of the
kinetic energy operator.
\be
\mu_0(nL)  = \frac{1}{2} \, \bar{E}_{\rm kin} (nL)
\label{eq.41}
\ee
In Table \ref{tab.01} the values of $\mu_0(nL)$ are given  for different sets of
the parameters of the static potential $V_0(r)$.  
\begin{table}
\caption{The dynamical masses $\mu_0(nL)$ (in MeV) for
different light mesons (the current mass $m=0$).\label{tab.01}}
\begin{center}
\begin{tabular}{|l|l|l|l|l|l|l|l|} 
\hline
& $1S$\,& $2S$\,& $3S$\,& $4S$\,& $1P$\, & $2P$\, & $1D$\,  \\  \hline
Set A &    & &&&&& \\
$\sigma$=0.143 GeV$^2$\, & 298\,& 445\, & 557\, & 650\, & 399\, & 516\, & 480\, \\
$\alpha_{\rm eff}=0\,$  &  &&&&&&\\ \hline
Set B &    & &&&&& \\
$\sigma$=0.143 GeV$^2$\, & 375\,& 513\, & 616\, & 703\, & 436\, & 551\, & 508\, \\
$\alpha_{\rm eff}=0.45\,$  &  &&&&&&\\  \hline
Set C &    & &&&&& \\
$\sigma$=0.18 GeV$^2$\, & 335\,& 500\, & 625\, & 729\, & 448\, & 579\, & 539\, \\
$\alpha_{\rm eff}=0$\, & & &&&&& \\ 
\hline
\end{tabular}
\end{center}
\end{table}
From Table~\ref{tab.01} one can see that the influence of the Coulomb
interaction is rather weak even for an $\alpha_{\rm eff}$ as large as
$\alpha_{\rm eff}=0.45$, except for the $1S$ case, where 
it changes the dynamical mass by roughly 25\%.  This happens
because the sizes of the light mesons are large, e.g. the
root--mean--square radii $R(nL)$ for the different states are as follows:  
\[ R(1S) = 0.8\div 0.9\,\mbox{fm}; \quad R(2S) = 1.3\div
1.4\,\mbox{fm}; \quad R(3S)=1.6 \div 1.8\,\mbox{fm}; 
\] 
\[
 R(4S) = 1.9 \div 2.1\,\mbox{fm}; \quad R(1P) = 1.0\div 1.2\,\mbox{fm};
\quad R(2P) = 1.4\div 1.6\,\mbox{fm}; \] \begin{equation}
 R(1D)=1.3   \div 1.4\,\mbox{fm}; \quad R(2D) = 1.6 \div 1.8\,\mbox{fm}
\label{eq.42} 
\end{equation} 
At such long distances the Coulomb interaction is small, only $\la
10\%$ compared to the linear term $\sigma r$. Moreover one can not
exclude that at $r \ga 1.2$ fm screening of the Coulomb interaction may
be important and therefore the Coulomb term in the static potential is
even smaller and can be neglected, being important only for the $1S$
ground state.

To illustrate our results, the spin-averaged masses of the low-lying
mesons are presented in Table~\ref{tab.02} and compared to the experimental
values (isovector and isoscalar mesons) and also to  the masses from the
paper by Godfrey and Isgur \cite{ref.17}, where the same Salpeter equation is
solved for a different set of parameters:
\be
\sigma = 0.18 \, \mbox{GeV}^2, \quad \alpha_{\rm GI}(r) \leq \alpha_{cr} = 0.60,
\quad C_0 = -253, \, \mbox{MeV}  \quad  m=220 \, \mbox{MeV}.
\label{eq.43}
\ee
As seen from Eq.~(\ref{eq.43}) in \cite{ref.17} a rather large value was taken
for the current mass $m$ of a light quark, while in our calculations the best 
fit was obtained with Set A:
\be
\sigma = 0.143 \, \mbox{GeV}^2, \quad \alpha_{\rm eff} = 0, \quad m=0, \quad
 C_0 = -357 \,\mbox{MeV}.
\label{eq.44}
\ee
The constant $C_0$ in Eq.~(\ref{eq.44}) was chosen to fit 
$M_{\rm cog}(2^3S_J)=1424$ MeV.
\begin{table}
\caption{ The spin-averaged masses $M_{\rm cog}(nL)$ (in MeV) of the low-lying 
light mesons. \label{tab.02}}
\begin{center}
\begin{tabular}{|l|l|c|l|l|}
\hline
state & $2S$ & $3S$ & $1P$ & $2P$  \\ \hline
This paper  &&&& \\
$\sigma=0.143$ GeV$^2$      &  1424\, & 1870\, & 1241\, & 1707\, \\
$\alpha_{\rm eff}=0$ && && \\
$C_0=-357$ MeV & fit &
 & &    \\ \hline
 Ref.~\cite{ref.17} &  1420\, & 1970\, & 1260\, & 1820\,  \\ \hline
experiment &
$ 1424$\, & $>1800$\,& 1252$^{a)}\,$ & $1632^{c)}$\, \\
$(I=1)$  &$\pm 44$\, &&1306$^{b)}$\, & $1683^{d)}$ \,
 \\ \hline
\end{tabular}
\end{center}

$^{a)}$ This value of $M_{\rm cog}(1P)$ is obtained if $a_0(980)$ belongs to the
$1^3P_J$ multiplet.

$^{b)}$ This value of $M_{\rm cog}(1P)$ is obtained if $a_0(1450)$ belongs to
the $1^3P_J$ multiplet.

$^{c)}$ This value of $M_{\rm cog}(2^3P_J)$  is obtained if $a_2(1660)$ belongs
to the $2^3P_J$  multiplet.

$^{d)}$ This value of $M_{\rm cog}(2^3P_J)$  corresponds to the case when
$a_2(1750)$   belongs to the $2^3P_J$ multiplet.
\end{table}

In Table~\ref{tab.02} the experimental numbers refer to the isovector mesons
which are not mixed with $s\bar{s}$ and are expected not to have a
large hadronic shift. From this table one can see that

(i) a better agreement with the experimental  masses is obtained if
$a_0(980)$ is a member of the $1^3P_J$ multiplet;

(ii) in our calculations the masses of the $3S$ and $2P$ states lie about 100 MeV
lower than in \cite{ref.17} and are closer to the experimental numbers for
$M_{\rm cog}(2 a_J$) and $\pi(1800).$

With the use of the dynamical masses $\mu_0(nL)=m_q$, presented in
Table~\ref{tab.01}, the nonperturbative HF splitting can be calculated, 
since from Eq.~(\ref{eq.15}) we obtain
\be
\Delta^{\rm NP}_{\rm HF}(nL) = \frac{2d}{\mu^2_0(nL)} \; 
\left(J_1 + \frac{d_1}{d}J_2 \right),
\label{eq.45}
\ee
where we have taken into account the second correlator $D_1(x)$ in
Eq.~(4) to have the opportunity to vary the values of the correlation
length $T_g$. In particular for $T_g=0.2$ fm the ratio
$d_1/d\approx 1/3$ was found in Ref.~\cite{ref.11}.

In Eq.~(\ref{eq.45})
\be
J_1 = \langle r \, K_1 (r/T_g) \rangle_{nL}, \quad
J_2 = \langle r \, K_1 (r/T_g) \rangle - \frac{1}{3T_g}
\langle r^2 K_0 \,(\frac{r}{T_g}) \rangle
\label{eq.46}
\ee
Here it is assumed that the gluonic correlation lengths $T_g$  and
$T_g^{(1)}$ in Eq.~(\ref{eq.15}) are equal as it was observed in
lattice measurements of $D(x)$  and $D_1(x)$  for $n_f=0$
\cite{ref.11,ref.13}.  We shall also fix the string tension $\sigma$
and from the definition Eq.~(\ref{eq.22}) the parameter $d$ is
\begin{equation}
d = \frac{\sigma}{\pi T_g^2} .
\label{eq.46a}
\end{equation} 
We estimate the accuracy of the calculated numbers to be about 10\%.
The nonperturbative HF splittings of the $S$-wave and $P$-wave  light
mesons are given in Table \ref{tab.03} for two values of the correlation length:
$T_g=0.5$ fm and $T_g=0.2$ fm 
(in both cases $\sigma=0.143$ GeV$^2$, $\alpha_{\rm eff}=0$).

\begin{table}
\caption{The nonperturbative HF splittings $\Delta^{\rm NP}_{\rm HF}(nL)$ 
(in MeV)for light mesons. \label{tab.03}}
\begin{center}
\begin{tabular}{|l|l|l|l|l|l|}
\hline
 state & $1S$ & $2S$ & $3S$ & $1P$ & $2P$ \\ \hline
 $T_g=0.3$ fm & 125 & 56 & 30 & 44 & 27 \\ \hline
 $T_g=0.2$ fm & 96 & 48 & 25 & 24 & 20 \\ \hline
\end{tabular}
\end{center}
\end{table}
As seen from Table~\ref{tab.03} the nonperturbative HF shift is large, $\approx
100$ MeV, for the $1S$ ground state; for other states the numbers weakly
depend on the value of $T_g$ with the exception of the $1P$ state for
which $\Delta_{\rm HF}^{\rm NP}$  is different for $T_g\approx 0.3$ fm
and $T_g\approx 0.2$ fm, which are taken from the lattice measurements
of the gluonic correlators \cite{ref.11}-\cite{ref.12}.  In most cases
the magnitude of HF splitting is between 20$\div 50$  MeV.

We consider also the $P$-wave mesons composed of a strange quark and
antiquark taking for the current mass of a strange quark $m_s=170$
MeV. Then the dynamical mass of the $s$ quark for different $nL$  states
turns out to be about 50 MeV higher than for a light quark (cf
Table \ref{tab.01}), in particular
\be
\mu_0(2S, s\bar{s}) = 505 \, \mbox{MeV}, \quad \mu_0(1P, s\bar{s})  =
454\,\mbox{MeV}, \quad \mu_0(2P,s\bar{s}) = 566\,\mbox{MeV} .
\ee
Correspondingly, the spin-averaged masses of the $s\bar{s}$ mesons
appear to be about 170 MeV higher than those for light mesons, e.g.
taking the  set $A$ of the parameters (\ref{eq.32}) and the constant
$C_0=-250$  MeV, defined from a fit to the spin-averaged mass of the $2S$
states ($\phi$(1680) and $\eta(1440)$), we have obtained that
\be
M_{\rm cog}(1P,s\bar{s}) = 1424\,\mbox{MeV},\, M_{\rm cog}(2P,s\bar{s}) =
1885\,\mbox{MeV} .
\ee
At this point it is of interest to note that $M_{\rm
cog}(1P,s\bar{s})$ coincides with the center of gravity of the
multiplet: $f_0(1370)$, $f_1(1420)$, $f_2(1430)$  which  are expected
to have a large $s\bar{s}$ admixture, but it is 50 MeV smaller if
$f_2(2P,s\bar{s})$ is identified with the $f'_2(1525)$ meson.

For the $1P\, s\bar{s}$ state the nonperturbative HF shift can be calculated 
from the expression (\ref{eq.25})  for $T_g=0.3$  fm and Eq.~(\ref{eq.45}) for
$T_g=0.2$ fm with the following result:
\be
\Delta^{\rm NP}_{\rm HF}(1P,s\bar{s}) =
\left\{
\begin{array}{ll}
37\,\mbox{MeV}, & T_g = 0.3 \,\mbox{fm} , \\
\\
20\,\mbox{MeV}, & T_g = 0.2 \,\mbox{fm} .
\end{array}
\right.
\ee

\section{Perturbative hyperfine splittings}

From experiment it is known that the HF and  fine structure splittings are 
practically small for all light mesons (with the exception of the $\pi$ and $K$
mesons) small compared to their masses and therefore the
spin-dependent effects can be considered as a perturbation. Then, as
was shown in Ref.~\cite{ref.08}, the spin dependent potentials can be derived by
averaging the spin factors which are present inside the meson Green's
function defined in a gauge invariant way. In this approach the expansion
in  inverse quark masses is not used and in Ref.~\cite{ref.08}  it was deduced
that to order $\alpha_s$ all perturbative spin-dependent potentials
$V_i(r)$ ($i=1,2,3,4$) for light mesons coincide with those in heavy
quarkonia with the only difference that the pole mass of a quark should
be repaced by the dynamical mass $\mu_0(nL)$ of a light quark (for a
heavy quark $\mu_0(nL)$  coincides with the current mass to order $\alpha_s$). 
In particular, the perturbative spin-spin potential between a light quark and 
a light antiquark is defined as
\be
V^{\rm P}_{\rm HF} (r) = \frac{V^{\rm P}_4(r)}{3\mu^2_0(nL)} .
\label{eq.48}
\ee
Then for the $S$-wave mesons the perturbative HF splitting is given by
the well-known expression:
\be
\Delta^{\rm P}_{\rm HF} (nS) =
\frac{8}{9}\,\frac{\alpha_s(\mu)}{\mu^2_0(nS)} \mid
R_{n 0}(0)\mid^2,
\label{eq.49}
\ee
where $\alpha_s(\mu)$ is the strong coupling in the $\overline{MS}$
renormalization scheme. 
In Ref.~\cite{ref.17} the spin-spin interaction was modified with a
smearing function with a characteristic momentum scale of about 1.8 GeV.
Consequently we can write in Eq.~(\ref{eq.49}) for the $S$-wave mesons
\be
\alpha_s(\mu)\approx \alpha_s(1.8\,\mbox{MeV})=
\alpha_s(M_{\tau})\approx 0.31 \div 0.33 .
\label{eq.50}
\ee
Since the scale $\mu$  coincides with the mass $M_{\tau}$ of the
$\tau$-lepton we take here $\alpha_s(\mu)=0.31$.

The wave function at the  origin entering Eq.~(\ref{eq.49}) cannot be
precisely defined for the Salpeter equation, since the expansion of the
wave function $\psi_{nL}(r)$ (18) in a basis (which is used here
for the numerical calculations as suggested in
Ref.~\cite{ref.19}) is diverging at the point $r=0$. Therefore  we
define $R_{n0}(0)\equiv \psi(nS,r=0)$ as in the einbein approach~\cite{ref.08} 
taking also into account the Coulomb interaction which
gives a correction of about $10\div 20\%$ and the largest one is for the
ground state $(\approx 30\%)$. Then $ R_{n0}(0)$  can be presented in
the form
\be
R_{n0}(0) = \sqrt{\mu_0(nS)\sigma} \,\xi(nS) ,
\label{eq.51}
\ee
where the coefficients $\xi(nS)$ are the following: ($\alpha_{\rm eff}=0.39$),
$\xi(1S)=1.31$, $\xi(2S)=1.20$, $\xi(3S)=1.16$,
$\xi(4S)=1.14$ and the values of the wave function at the origin are
\begin{eqnarray}
R_{10}(0) & = & 0.294\,\mbox{GeV}^{3/2}, \quad R_{20}(0) = 0.30\,\mbox{GeV}^{3/2},
\nonumber \\
R_{30}(0) & = & 0.325\,\mbox{GeV}^{3/2}, \quad R_{40}(0) = 0.34\,\mbox{GeV}^{3/2}.
\label{eq.52}
\end{eqnarray}
From these numbers one can see that the wave function at the origin is almost
constant, but slowly growing because of the increase of the dynamical
mass $\mu_0(nS)$ with $n$.

The values of the perturbative splittings for the $nS$  states are given in
Table \ref{tab.04} $(\alpha_{\overline{MS}}=\alpha_s=0.31)$. If one neglects the
Coulomb correction in the wave function $R_{n0}(0)$  then $\Delta^{\rm P}_{\rm
HF}$ will be about 30-50\% smaller. To check our choice of $R_{n0}(0)$
one can calculate the leptonic width of $\rho(770)$:
\be
\Gamma_{e^+e^-} = \frac{2\alpha^2\mid R_{10}(0)\mid^2}{M^2_{\rho}} \Biggl
(1-\frac{16}{3\pi}\alpha_s \Biggr ),
\label{eq.53}
\ee
which gives the following value for the leptonic width
($\alpha_{\overline{MS}}=0.31;\,\alpha=1/137$)
\be
\Gamma_{e^+e^-} (\rho(770))  = 7.36 \,\mbox{keV} ,
\label{eq.54}
\ee
that turns out to be in good agreement with the experimental number
$\Gamma_{e^+e^-}({\rm exp})=6.77 \pm 0.32$  keV \cite{ref.08}  (for
$\alpha_{\overline{MS}}=0.33$ the leptonic width is $\Gamma_{e^+e^-}=6.8$ keV).

From the number (\ref{eq.52}) for $R_{20}$  one can expect that
$\Gamma_{e^+e^-}(\rho(1450))\approx 1.7$ keV and the fraction
$\Gamma_{e^+e^-}/\Gamma_{\rm total}$  for $\rho(1450)$ is seven times
smaller than for $\rho(770)$.
\begin{table}
\caption{The hyperfine splittings of the $S$-wave light mesons (in MeV) 
with $\alpha_{\overline{MS}}=0.31$ \label{tab.04}}
\begin{center}
\begin{tabular}{|l|l|c|l|c|}
\hline
& $1S$ & $2S$ & $3S$ & $4S$\\ \hline
$\Delta^{\rm P}_{\rm HF}$ & 194 & 125 & 94 & 75(60)\\ 
\hline
$\Delta_{\rm HF}({\rm total})$,\, $T_g=0.3$ fm & 329 & 185 & 144 & 96\\
\hline
$\Delta_{\rm HF}({\rm total})$,\, $T_g=0.2$ fm & 290 & 173 & 119 & 95\\
 \hline
experiment & & $165\pm 100$ && \\ 
\hline
\end{tabular}
\end{center}
\end{table}

From the comparison of the nonperturbative and perturbative spin-spin
splittings in Tables \ref{tab.03} and \ref{tab.04} one can see that for
all $nS$-states $(n\not= 1)$ the perturbative splitting $\Delta^{\rm P}_{\rm
HF}(nS)$   turns out to be about two times larger than    $\Delta^{\rm
NP}_{\rm HF}$  while for  the $1S$ state the nonperturbative contribution
is larger. It is  about 60\% of $\Delta^{\rm P}_{\rm HF}(1S)$.

Knowing the HF splittings  we can  calculate the masses of  the isovector
mesons (see Table~\ref{tab.05}), neglecting the coupling to the other channels.
\begin{table}
\caption{The predicted masses of the $S$-wave mesons in MeV
 ($T_g=0.2$ fm). \label{tab.05}}
\begin{center}
\begin{tabular}{|l|c|c|c|c|c|c|}\hline
& $\pi(2S)$ &  $\rho(2S)$ & $\pi(3S)$ & $\rho(3S)$ & $\pi(4S)$ &
$\rho(4S)$ \\ \hline
theory  & 1294 & 1467 & 1781 & 1900 & 2170 & 2265 \\ \hline
Ref.~\cite{ref.17} & 1300 & 1450 & 1880 & 2000 & -- & --\\ \hline
experiment  & $1300\pm 100$ & $1465\pm$25 & 1800$\pm$13 & -- &
-- & 2149$\pm 17$  $^{a)}$ \\ \hline
\end{tabular}
\end{center}

\vspace{3mm}

$^{a)}$ The mixing of $4^3S_1$ and $2^3D_1$ states is not taken into account.
\end{table}

We would like to notice here  that all our calculations were done for
a massless quark (antiquark)  with only two parameters: the string
tension $\sigma=0.143$ GeV$^2$ (which defines the dynamical mass  of
the quark (antiquark) $\mu_0(nS)$ and the spin-averaged spectrum)  and
the value $\alpha_{\overline{MS}}\approx
\alpha_{\overline{MS}}(M_{\tau})\approx 0.31$ suggesting  that the
characteristic ``smearing radius'' is small as in
Ref.~\cite{ref.17}. Still, in such a simple picture the agreement with
experiment is reasonably good and our masses for the $3S$ states are about 100
MeV lower than in Ref.~\cite{ref.17} and close to the experimental mass
of $\pi(1800)$.

To obtain the masses of  the $4S$ states one needs to take into account the
mixing of these states with the $2D$ states with $M_{\rm cog}(2D)=1972$ MeV
(for the same set of parameters A).  The mixing will be done elsewhere.

\section{The masses of the \lowercase{$b_1$} and \lowercase{$h_1$} mesons}

For the $P$-wave state the perturbative  HF splitting is of order
$\alpha^2_s$ and is expected to be small. To estimate the
perturbative contribution  one can use the expression ~\cite{ref.20}
\be
\Delta^{\rm P}_{\rm HF}  = \frac{8}{9}\,\frac{\alpha^2_{\overline{MS}}}{\pi m^2_q}
\,\Biggl [\frac{1}{4}-\frac{1}{3} n_f \Biggr ] \langle r^{-3} \rangle_{nP}
\to
 \frac{2}{3}\,\frac{\alpha^2_{\overline{MS}}}{\pi \mu^2_0(nP)}
\langle r^{-3} \rangle_{nP}, \quad (n_f=3) .
\label{eq.55}
\ee
This perturbative HF shift is negative and in (\ref{eq.55}) $m_q$  is
repaced by the dynamical mass of a light quark. This is allowed since
the $P$-wave HF potential $V^{\rm P}_4(r)$ does neither depend on
the renormalization scale  nor on the mass of a quark (antiquark). This
expression follows from the perturbative spin-spin potential for
$L\not= 0$ \cite{ref.21}
\begin{eqnarray}
V^{\rm P}_{\rm HF} (r) & = & \frac{1}{3m^2_q} V^{\rm P}_4(r), \nonumber \\
V^{\rm P}_4(r) & = & \frac{8}{3\pi}
\,\alpha^2_{\overline{MS}}  \left(\frac{1}{3}n_f -\frac{1}{4} \right)
\nabla^2\frac{\log\,r}{r}=
= \frac{8}{3\pi} \alpha^2_{\overline{MS}}
 \left (\frac{1}{4} -\frac{1}{3}n_f \right) \,\frac{1}{r^3}.
\label{eq.56}
\end{eqnarray}
This short-range spin-spin potential has a characteristic size $R_{\rm
HF}$ which can be estimated from the value of the matrix element
$\langle r^{-3}\rangle_{nP}$:
\be
\langle r^{-3}\rangle_{1P}  = 0.019 \,\mbox{GeV}^3, \quad
\langle r^{-3}\rangle_{2P} = 0.030 \,\mbox{GeV}^3.
\label{eq.57}
\ee
If $R_{\rm HF}(nP) =(\langle r^{-3}\rangle_{nP})^{-1/3}$  then $R_{\rm
HF}(1P)\approx 0.75$  fm and  $R_{\rm HF}(2P)\approx 0.65$ fm are rather
large. From these estimates one can conclude that for the $P$-wave
states $R_{\rm HF} (nP)\approx 0.65$  fm appears to be much larger
than for the $nS$-states where in the smearing function $R_{\rm
HF}(nS)=(1.8$ GeV)$^{-1}\approx 0.11$ fm was taken from Ref.~\cite{ref.17}  At
the distances  $R_{\rm HF}\approx 0.65$ fm the value of
$\alpha_{\overline{MS}}$ needs to be taken at the smaller
renormalization scale and is  very close to the freezing value
$\alpha_{\overline{MS}}(q=0)$  which is expected to be
$\alpha_{\overline{MS}}(q=0)\approx 0.5$. Therefore    here we take
$\alpha_{\overline{MS}}(q=0)\approx 0.45$. The numbers obtained
from Eq.~(\ref{eq.55})
\be
\Delta^{\rm P}_{\rm HF} (1P) = -5.1 \,\mbox{MeV},\quad
\Delta^{\rm P}_{\rm HF} (2P) = -4.8 \,\mbox{MeV},
\label{eq.58}
\ee
are much smaller than the nonperturbative shift given in Table \ref{tab.03} and
have opposite signs. Combining both contributions, one obtains the total
HF splitting,
\be
\Delta_{\rm HF}(1P) = \left\{
\begin{array}{ll}
 39 \,\mbox{MeV}, & \quad \mbox{if}\; T_g = 0.3\,\mbox{fm}\\
 \\
 19\,\mbox{MeV}, & \quad \mbox{if}\; T_g = 0.2\,\mbox{fm}
\end{array}
\right.
\label{eq.59}
\ee
or the average number $\Delta_{\rm HF} = 29 \pm 10$ MeV. 
Knowing  the mass of $b_1(1235)$, 
\be
M(b_1(1P))  = 1229.5\pm 3.2\,\mbox{MeV},
\label{eq.66a}
\ee
the predicted mass for the center of gravity of the $1^3P_J$ multiplet
($T_g=0.3$ fm) is
\be
M_{\rm cog}(1^3P_J) = 1258 \pm 3.2\, ({\rm exp}) \pm 10\, ({\rm th})\,\mbox{MeV}.
\label{eq.66b}
\ee
The number obtained for $M_{\rm cog}(1^3P_J)$  is in surprisingly good
agreement  with the experimental  mass $M_{\rm cog}(1^3P_J,{\rm
exp})=1252$ MeV, if $a_0(980)$ belongs to the $1^3P_J$ multiplet,  and
does not agree with $M_{\rm cog}(1^3P_J)=1306$   MeV  obtained in the
case that $a_0(1450)$  belongs to the $1^3P_J$  multiplet.  Thus a
strong correlation between the masses  of $M_{\rm cog}(1{}^3 P_J)$  and
$b_1(1235)$ follows from our analysis and to fit the experimental data
one must assume that $a_0(980)$ belongs to the $1^3P_J$ multiplet and
is a $q\bar{q}$ state.

Then $a_0(1450)$ can be considered as a member of the $2^3P_J$
multiplet with $M_{\rm cog}(2P)=1633$ MeV from Table~\ref{tab.02} and therefore
with the use of the total HF shift  we predict for the mass of
$b_1(2P)$:
\be
M(b_1(2P))  = 1610 \div 1618 \,\mbox{MeV} ,
\label{eq.62}
\ee
since the \underline{total} HF shift from Table \ref{tab.03} 
and Eq.~(\ref{eq.58}) is
\be
\Delta_{\rm HF}(2P) = \left \{
\begin{array}{ll}
 22\,\mbox{MeV}\; T_g=0.3\,\mbox{fm}), \\
 \\
 15\,\mbox{MeV}\; T_g=0.2\,\mbox{fm}) .
\end{array}
\right.
\label{eq.63}
\ee
In the approximation of closed channels used here the HF shift of
$h_1(1170)$ and  $b_1(1235)$   should be the  same, see  Eq.~(\ref{eq.63}).
However, for $h_1(1170)$ the experimental value of the  HF shift is
larger, $73 \pm 19$ MeV, and therefore one cannot exclude that
$h_1(1170)$ has a small hadronic  shift, $\Delta M_{\rm had}=35 \pm 20$ MeV
(note that $h_1(1170)$ has a much larger width, $\Gamma(h_1)\approx 360$
MeV, than $b_1(1235)$).  There exists also the state $h_1(1380)$ with
$M(^{1}P_1)=1386\pm 19$ MeV.  It is assumed that $h_1(1380)$  is mostly
composed of a strange quark and antiquark.  Then from
the calculated $\Delta_{\rm HF}({\rm total})\approx 35$ MeV 
($T_g = 0.3$ fm and $\Delta^{\rm P}_{\rm HF} = 4$ MeV) one can obtain
the center of gravity of the $1^3P_J$  multiplet of $s\bar{s}$
mesons:
\be
M_{\rm cog}(1^3P_J,\, s\bar{s})
\approx M(1^1P_1) + 35 \,\mbox{MeV} \approx 1425 \pm 19\,\mbox{MeV}.
\label{eq.64}
\ee
This number can be compared with $M_{\rm cog}(1^3P_J)$ obtained in
the case if $f_0(1370)$, $f_1(1426)$, $f_2(1430)$ are members of
the $1^3P_J$  multiplet and mostly $s\bar{s}$ states:
\be
M^{(1)}_{\rm cog} (1^3P_J) \approx 1422\,\mbox{MeV}
\label{eq.65}
\ee
and this experimental mass is in good agreement with the predicted
mass (\ref{eq.64}).  In the other case, when $f_2(1525)$ is a member of the
$1^3P_J$  multiplet, the ``experimental'' value of the center of gravity,
\be
M^{(2)}_{\rm cog} (2^3P_J) \approx 1474\,\mbox{MeV}
\label{eq.66}
\ee
is not correlated with the mass of $h_1(1380)$ and the shift of the mass
of $h_1(1380)$ appears to be larger (about 80 MeV) than in our
calculations.

\section{Conclusions}

We investigated the nonperturbative spin-spin interaction in light
mesons and established that

1. For the $1S$ state the HF shift due to the nonperturbative effects is
rather large, because the dynamical mass is relatively small, 
so that $\Delta^{\rm NP}_{\rm HF}\approx 0.4$
$\Delta_{\rm HF}$($1S$, total), while for the excited $nS$ states it is
only about 15\% of the total shift.

2. Due to the positive sign of the nonperturbative HF splitting the mass of the
$n^1P_1$  state is strongly correlated with $M_{\rm cog}(n^3P_J)$
being $30\pm 10$ MeV smaller than $M_{\rm cog}(n^3P_J)$. The value of this
shift depends on the gluonic correlation length adopted.

3. With the use of the mass of $b_1(1235)$  our predicted mass of
$M_{\rm cog}(1^3P_J, \, I=1)$  is $1258\pm 10$ MeV and this number is in
agreement with the experimental masses of the $a_J(1P)$  mesons only if
$a_0(980)$ belongs to the $1^3P_J$  multiplet.

4. For $b_1(2P)$  we predict the mass $M(b_1(2P))\approx 1.62$  GeV.

5. Our analysis can be applied also to the isoscalar mesons where
$h_1(1170)$ and $M_{\rm cog}(1^3P_J)=1245$  MeV lie rather close to
each other if $f_0(980)$ is a member of the $1^3P_J$ multiplet.

6. In the approximation when $h_1(1380)$, $f_0(1370),\, f_1(1420)$,
$f_2(1430)$ are considered to be composed mainly of a strange quark and
antiquark, the difference $\Delta=M_{\rm cog}(1^3P_J,s\bar{s})$
$-M(h_1(1380))\approx 35$  MeV is in full agreement with our estimate
of the  nonperturbative HF shift, $\Delta^{\rm NP}_{\rm HF}\approx 35$
MeV for the correlation  length $T_g=0.3$ fm.

7. The preferable value of the gluonic correlation length $T_g=0.3$ fm
was obtained from our analysis of the HF splittings of different mesons
in accordance with the lattice data of Ref~\cite{ref.12}.

\vskip 10mm

This paper was partly supported by the grant RFFI-00-02-17836.
\end{document}